\documentstyle[12pt,aps]{revtex}
\input epsf
\begin{document}
\title{\hfill\parbox[t]{2in}{\rm\small\baselineskip 10pt
{~~~~~~JLAB-THY-98-03*}\vfill~}
\vskip 2cm
Duality-Violating $1/m_Q$ Effects in Heavy Quark Decay}

\vskip 1.0cm

\author{Nathan Isgur}
\address{Jefferson Lab\\
12000 Jefferson Avenue,
Newport News, Virginia 23606}
\maketitle

\vspace{3.0 cm}
\begin{center}  {\bf Abstract}  \end{center}
\vspace{.4 cm}
\begin{abstract}

	I identify a source of $\Lambda_{QCD}/m_Q$ corrections to the assumption of 
quark-hadron duality in the application of heavy quark methods to inclusive
heavy quark decays.  These corrections 
could substantially affect the accuracy of such methods in practical
applications and in particular
compromise their utility for the extraction of the
Cabibbo-Kobayashi-Maskawa matrix element
$V_{cb}$.
 
\bigskip\bigskip\bigskip\bigskip\bigskip\bigskip
\noindent {*an abbreviated version of the original JLAB-THY-98-03 
entitled ``Duality in Inclusive Semileptonic Heavy Quark Decay"}

\end{abstract}

\newpage

  Although the classic application of heavy quark symmetry is in the exclusive 
semileptonic decays of heavy quarks \cite{IWoriginal}, there has also been substantial work on
using heavy quark effective theory (HQET) \cite{HQEToriginal} to systematically improve decay 
predictions for inclusive decays of heavy hadrons \cite{inclorig,Russianinclusives,otherinclusives}.  In these 
inclusive applications, decays
are treated in an operator product expansion (OPE)  which leads {\it via} HQET
to a $1/m_Q$ expansion in which the leading term is free quark decay and
$1/m_Q$ terms appear to be absent.  Although these calculations have become very 
sophisticated \cite{Russianinclusives,otherinclusives}, it is widely 
appreciated \cite{Russianinclusives,otherinclusives,georgicomm,Lipkin} that there remains a basic 
unproved hypothesis in their derivation:  the assumption of quark-hadron duality.
It is the accuracy of this assumption that I want to call into question here.

	 While supposedly of wide validity, recent applications have centered around the hope that this approach
offers an alternative to the classic exclusive methods
for determining $V_{cb}$, and I will accordingly focus most of my remarks
on the case $b \rightarrow c \ell \bar \nu_{\ell}$ where both quarks
are heavy.  In inclusive $b\rightarrow c \ell \bar \nu_{\ell}$ 
decays, which materialize as $\bar B\rightarrow X_c \ell \bar \nu_{\ell}$, about 65\%
of the $X_c$ spectrum is known to be due to the very narrow ground states $D$ and $D^*$.  The
relatively narrow $s_{\ell}^{\pi_{\ell}}={3 \over 2}^+$ states \cite{IWspec} 
$D_2^*(2460)$ and $D_1(2420)$ account for perhaps
another 5\% of the rate, and it may be assumed that the remaining rate involves decays 
to higher mass resonances (quarkonia and hydrids) and continua \cite{exlusincluscomment}.  The 
inclusive calculations predict continuous $X_c$ spectra which are assumed to be
dual to the true hadronic spectrum (see Fig. 1).

  A picture like Fig.~1  might lead one to dismiss the duality approximation
since the inclusive spectrum clearly does not meet the usual requirement
that it be far above the resonance region \cite{masscomment}.  {\it I.e.}, normally the accuracy of 
quark-hadron duality would be determined by a parameter $\Lambda_{QCD}/ E$ where 
the relevent energy scale $E$
is the
mean hadronic excitation energy $\Delta m_{X_c} \equiv \bar m_{X_c}-m_D$.  However, as first explained by Shifman and 
Voloshin \cite{SV,NIonSV}, this is {\it not}
the expansion variable in this case: duality for heavy-to-heavy semileptonic decays sets in {\it 
at threshold} since even as $\delta m \equiv m_b - m_c$ (and therefore $\Delta m_{X_c}$)
approaches zero,
as $m_b \rightarrow \infty$
the heavy recoiling $c$ quark has an energy much greater than $\Lambda_{QCD}$ so that it
{\it is} a free quark in leading order.  In the small velocity (SV) limit, 
it {\it must} therefore hadronize with
unit probability (up to potential $\Lambda_{QCD}/ m_Q$ corrections) as $D$ and $D^*$. 
This ``cannonball" approximation is in fact an essential part of the physical basis of 
the HQET expansion in $1/ m_Q$.  Thus the question is not whether duality holds in semileptonic
heavy quark decays,  but rather how accurately it holds.

\bigskip

%
%
\begin{center}
~
\epsfxsize=3.0in  \epsfbox{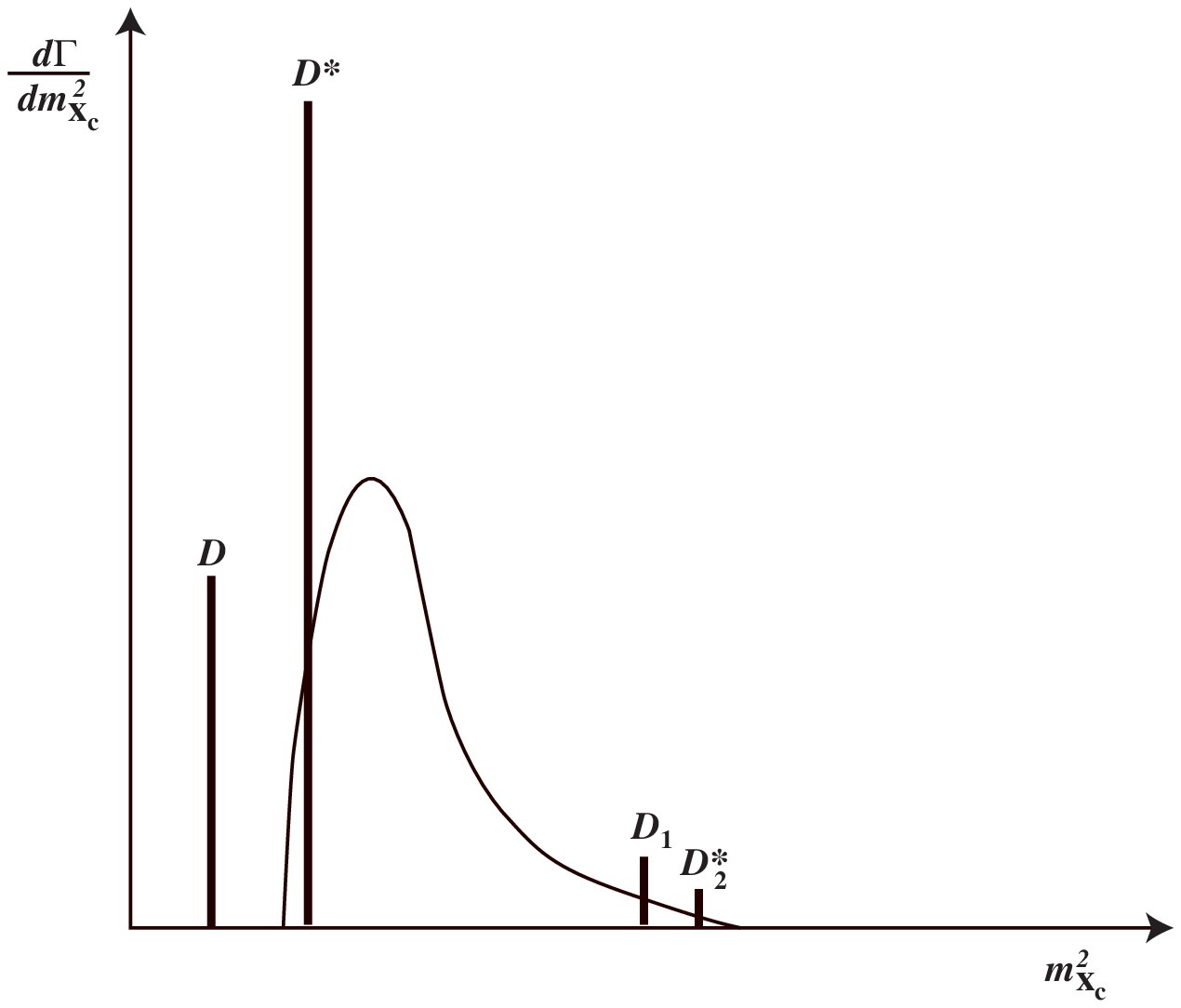}
\vspace*{0.1in}
~
\end{center}

\noindent{ Fig. 1: A sketch for $b \rightarrow c$ semileptonic decay of the continuous inclusive recoil 
spectrum of the OPE calculations (smooth curve) compared to the known hadronic spectrum (shown as
individual resonance lines).}

\bigskip\bigskip

    Up to {\it caveats} regarding the unknown accuracy of the assumption
of duality, the combined HQET and OPE methods 
indicate that inclusive calculations should in fact
be accurate up to corrections of order $\Lambda^2_{QCD}/
m^2_Q$.  Here I will identify a source of
duality-violation which leads to $\Lambda_{QCD}/m_Q$ corrections
for any finite final quark kinetic energy.  The problems are 
revealed by considering a
Bjorken sum rule \cite{BjSR} which may be viewed as an extension of  Shifman-Voloshin duality to arbitrary 
recoils.  Bjorken's sum rule guarantees that, as $m_b \rightarrow \infty$, duality will be enforced locally in the
semileptonic decay Dalitz plot of rate versus $w-1$ and $E_{\ell}$ (where $w \equiv v \cdot v'$ is the usual heavy 
quark double-velocity
variable and $E_{\ell}$ is the lepton energy).  For regions of the Dalitz plot for which $w-1$ is not
large (and in $b \rightarrow c$ decay nearly the whole Dalitz plot satisfies this condition), the Bjorken sum rule
explicitly relates the loss of total rate from the ``elastic" $s_{\ell}^{\pi_{\ell}}={1 \over 2}^-$ 
channels, as the Isgur-Wise function falls, to the
turn-on of the production of $s_{\ell}^{\pi_{\ell}}={1 \over 2}^+$ and ${3 \over 2}^+$
states \cite{IWonBj}.  

\bigskip\bigskip

%
%
\begin{center}
~
\epsfxsize=3.5in  \epsfbox{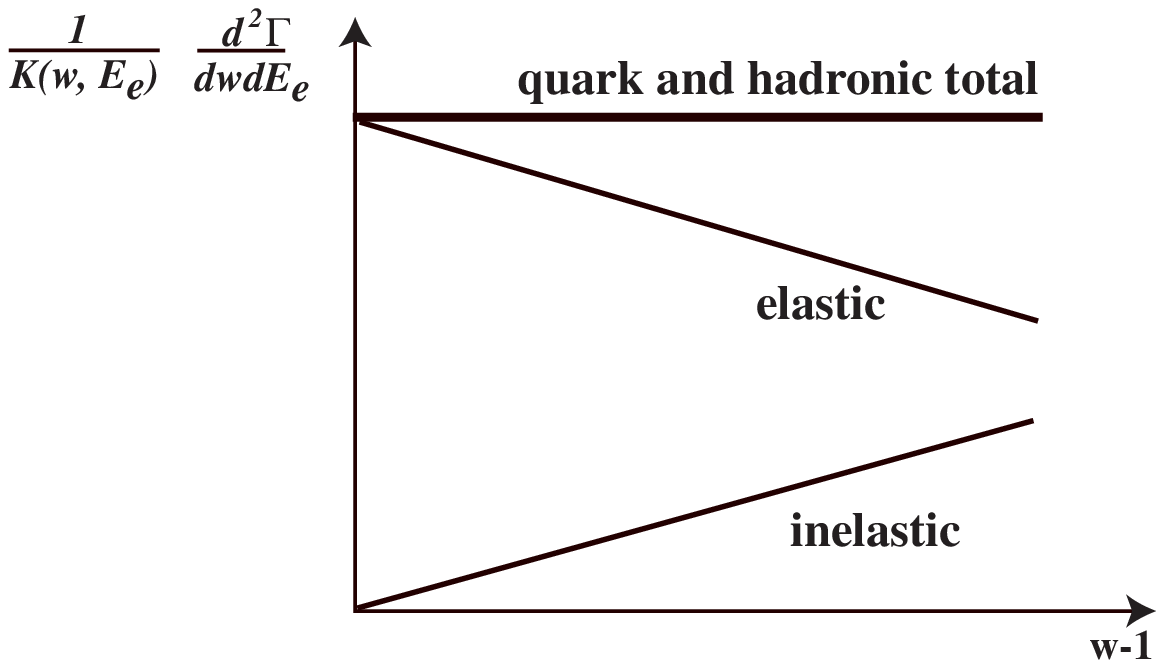}
\vspace*{0.1in}
~
\end{center}

\noindent{ Fig. 2:  The exact compensation by inelastic channels 
of the fall of the elastic rate 
in the linear region as $m_b \rightarrow \infty$.}

\bigskip\bigskip

   In particular, in this region the Isgur-Wise 
function may be taken to be linear:
\begin{equation}
\xi(w) \simeq 1-\rho^2(w-1) \equiv 1-\left[ {1\over 4}+\rho^2_{dyn} \right] (w-1)~~~,
\end{equation}
and if we define
\begin{equation}
{d^2\Gamma^{inclusive}_{quark} \over {dwdE_{\ell}}} = K(w,E_{\ell})
\end{equation}
then \cite{BjSR,IWonBj}, as $m_b \rightarrow \infty$,
\medskip
\begin{equation}
{d^2\Gamma^{inclusive}_{hadron} \over {dwdE_{\ell}}} = K(w,E_{\ell}) \Biggl( {w+1 \over 2} \vert \xi(w) \vert^2
+2(w-1) \left[ \sum_m \vert \tau^{(m)}_{1 \over 2}(1) \vert^2 
+2 \sum_p \vert \tau^{(p)}_{3 \over 2}(1) \vert^2 \right] \Biggr)
\end{equation}
up to corrections of order $(w-1)^2$.
With
\begin{equation}
\Biggl({w+1 \over 2}\Biggr) \vert \xi(w) \vert^2
\simeq 1- 2 \rho^2_{dyn}  (w-1)~~~,
\end{equation}
the Bjorken sum rule guarantees that for fixed $r \equiv m_c/m_b$, as $m_b \rightarrow \infty$ inelastic 
$s_{\ell}^{\pi_{\ell}}={1 \over 2}^+$ and ${3 \over 2}^+$ channels will
open up to give a semileptonic rate that exactly and locally compensate in the Dalitz plot the loss of
rate from the elastic channels due to $\rho^2_{dyn}$.
This situation is sketched in Figure 2; if it were
applicable to $b \rightarrow c$ decays, then quark-hadron duality would be exact.

   Having established  conditions for its validity as $m_b \rightarrow \infty$, 
it is easy to see why one should be concerned about 
quark-hadron duality for $b \rightarrow c$ decays.  
For fixed $r$,
$w-1$ lies in the fixed range from $0$ to $(1-r)^2/2r$, and as $m_b \rightarrow \infty$ any given hadronic threshold
collapses to the point $w=1$.  However, for finite $m_b$ there is a gap in $w-1$ in which the rate to the
elastic ${1 \over 2}^-$ channels falls by $\Lambda_{QCD}/m_Q$ terms 
but the potentially compensating excited state channels 
${1 \over 2}^+$ and ${3 \over 2}^+$ are not yet
kinematically allowed.  
More precisely, if $m_{D^{**}}$ is the mass of a generic charmed 
inelastic state, then $t^{**}_m=(m_B-m_{D^{**}})^2$  would be the threshold in $t$ for this state, 
corresponding to a value of $w-1$ 
in the quark-decay Dalitz plot of
\begin{equation}
   { {t_m-t^{**}_m} \over 2m_b m_c} \simeq (1-r) {\Delta \over m_c}
\end{equation}
where $t_m \equiv (m_B-m_D)^2 \simeq (m_b-m_c)^2$ and $\Delta \equiv m_{D^{**}}-m_D$.  Since $\Delta \simeq 500$ MeV 
and $(w-1)_{max} \simeq 0.6$, this region covers more than one third of the Dalitz plot and the compensation
is very substantially delayed:  see Figure 3.  Eqs. (5) and (1) show that this effect is of order
$\Lambda_{QCD}/m_Q$, seemingly at odds with the OPE result.

\bigskip

%
%
\begin{center}
~
\epsfxsize=3.1in  \epsfbox{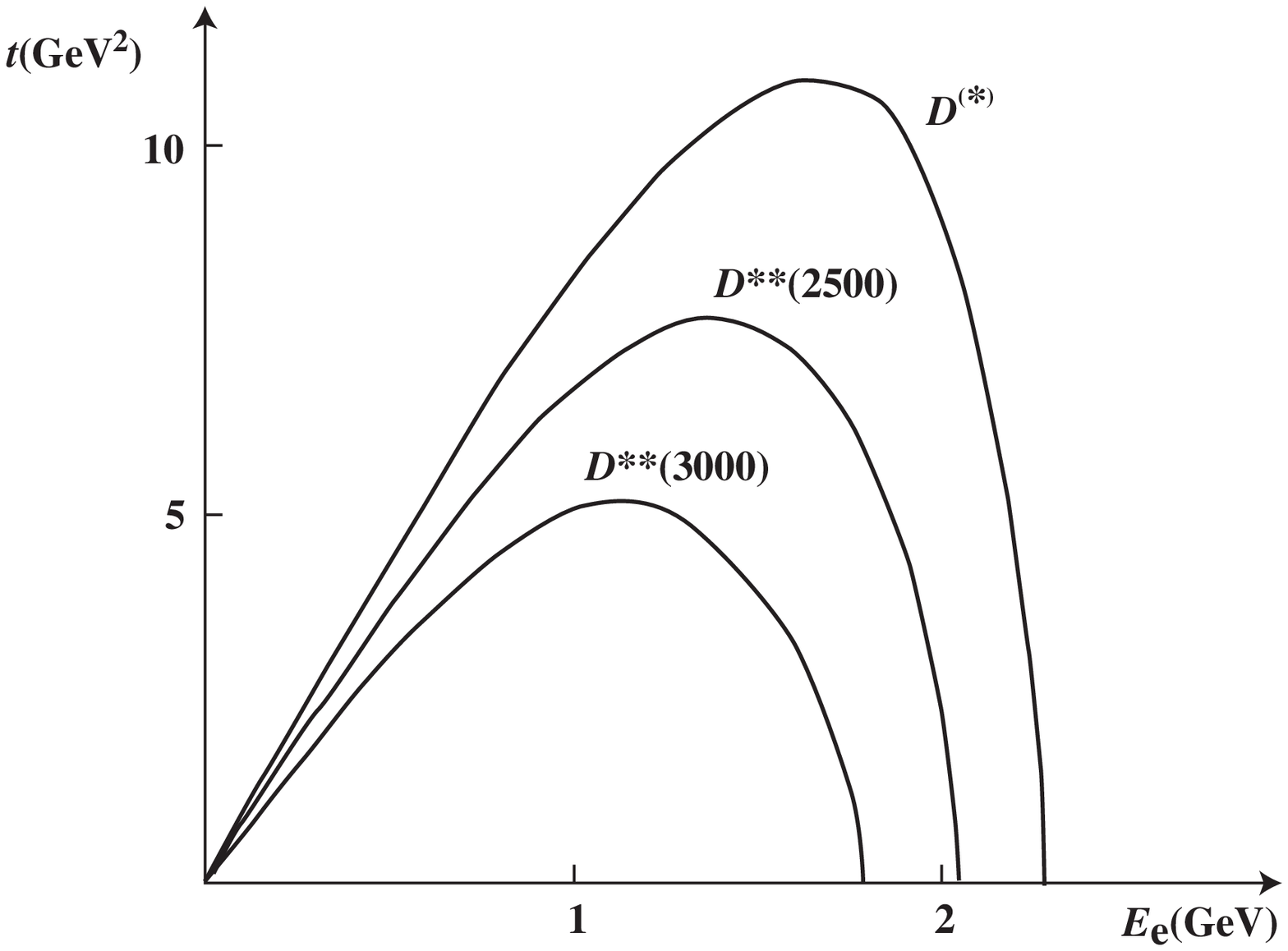}
\vspace*{0.1in}
~
\end{center}

\noindent{ Fig. 3: An overlay of the Dalitz plots for 
$\bar B\rightarrow D^{(*)} e \bar \nu_e$,
$\bar B \rightarrow D^{**}(2500) e \bar \nu_e$,
and $\bar B \rightarrow D^{**}(3000) e \bar \nu_e$. The $D^{(*)}$ mass is taken as
the hyperfine average of the  $D$ and $D^*$ masses; the two
$D^{**}$ masses are chosen for illustrative 
purposes.}

\bigskip

   Despite this apparent contradiction, there is actually no inconsistency: the OPE
result that the leading corrections to the inclusive rate are of order  $\Lambda^2_{QCD}/m^2_Q$
can still be valid as derived in the limit of large energy release in
the $b \rightarrow c$ transition, while $\Lambda_{QCD}/m_Q$
effects can arise for energy
releases of the order of $\Lambda_{QCD}$ due to 
a finite radius of convergence of the OPE. The main purpose of this paper is indeed
to call attention to this  effect.

   The basic issues can be most easily exposed by considering \cite{Russianinclusives}
spinless quarks coupled to a scalar field $\phi$ of mass $\mu$, and by studying
the decay $b \rightarrow c \phi$ with weak coupling constant $g$. 
Differential semileptonic decay rates have a more complex spin structure, but otherwise correspond to the case
$\mu=\sqrt{t}$; total semileptonic rates correspond to a weighted average
over kinematically allowed $\mu$ but, as we shall see below, this averaging
does not change the essentials of the problem.
In our simplified case
\begin{equation}
\Gamma (b \rightarrow c \phi) = {g^2p_{cb} \over 8 \pi m_b^2}
\end{equation}
where $p_{fi} \equiv [(m_i-m_f)^2-\mu^2]^{1/2}[(m_i+m_f)^2-\mu^2]^{1/2}/2m_i$
is the momentum of $\phi$ from the two-body decay of mass $m_i$ into masses $m_f$ and
$\mu$. 

    To compare Eq. (6) with a hadronic world 
(I initially consider a large $N_c$ world of narrow resonances, but will generalize below), define
\begin{equation}
\Gamma (B \rightarrow D^{(n)} \phi) = {g^2p_{D^{(n)} B} \over 8 \pi m_B^2}
\Bigl({{m_{D^{(n)}}m_B} \over m_cm_b}\Bigr) \vert \xi^{(n)}(\vec v_{D^{(n)} B})\vert^2
\end{equation}
where the generalized Isgur-Wise functions $\xi^{(n)}$ depend on
\begin{equation}
\vec v_{D^{(n)} B}=\vec p_{D^{(n)} B}/m_{D^{(n)}}~~,
\end{equation}
the recoil velocity of the $n^{th}$ excited state $D^{(n)}$ of the $D$ meson system,
and I have introduced some conventional mass factors to explicitly reflect hadronic normalizations. 
I next introduce a ``scaled energy release" variable
\begin{equation}
T^* \equiv  {m_b-m_c - \mu \over \Delta}~~,
\end{equation}
where $\Delta \equiv m_{D^{**}}-m_D$ is the mass gap to the first 
$s_{\ell}^{\pi_{\ell}}=1^-$ excited state of the $D$ meson system (corresponding to
$s_{\ell}^{\pi_{\ell}}={1 \over 2}^+$ and ${3 \over 2}^+$ in the physical
case where the $\bar d$ has $j^p={1 \over 2}^-$),
and an order
$\Lambda_{QCD}/m_Q$ expansion parameter $\epsilon$ defined by the expansion
\begin{equation}
\vert \xi_{DB} \vert^2 = 1 - \epsilon T^* + O(\epsilon^2)
\end{equation}
for small $T^*$ ({\it i.e.}, small charm quark velocities).
In the quark model \cite{ISGW,ISGW2} one would have
\begin{equation}
\epsilon  = {m_d (m_b-m_c) \over m_b m_c}
\end{equation}
where $m_d$ is the mass of the light spectator
antiquark $\bar d$ (or, more generally, of the ``brown muck").
Defining $\vert \xi_{D^{**}B} \vert^2 \equiv \sum_{m_{\ell}} \vert \xi^{m_{\ell}}_{D^{**}B} \vert^2$
(where $\xi^{m_{\ell}}_{D^{**}B}$ is the analog of $\xi_{DB}$ for transitions into the
lowest $s_{\ell}^{\pi_{\ell}}=1^-$ excited state with magnetic quantum number $m_{\ell}$), we would have
\begin{equation}
\vert \xi_{D^{**}B} \vert^2 = \epsilon (T^*-1) + O(\epsilon^2)
\end{equation}
from the (spinless) Bjorken sum rule
in the limit that it is saturated by the first $D^{**}$. Since in this limit
\begin{equation}
{p_{D^{(n)}B} \over p_{cb}} = \Bigl[ {m_{D^{(n)}} m_b \over m_c m_B} \Bigr]^{1/2} 
\Bigl({T^*-1 \over T^*} \Bigr)^{1/2}~~,
\end{equation}
we can obtain a model \cite{Russianinclusives} for 
\begin{equation} 
R \equiv {\sum_n \Gamma (B \rightarrow D^{(n)} \phi) \over \Gamma (b \rightarrow c \phi) }~~~.
\end{equation}
by truncating the sum over $n$ after the first $D^{**}$:
\begin{equation} 
R_1^{D^{**}} \equiv {\Gamma (B \rightarrow D \phi)+\Gamma (B \rightarrow D^{**} \phi) 
\over \Gamma (b \rightarrow c \phi) }
\end{equation}
\begin{equation} 
~~~~~~~=  [1 +{3\over 2} \epsilon - \epsilon T^*] \theta (T^*)
+ \epsilon {(T^*-1)^{3/2} \over {T^*}^{1/2}} \theta (T^*-1)~~~,
\end{equation}
wherein I have shown explicitly the two thresholds at $T^*=0$ and $T^*=1$.
It is interesting to observe that the quark model of Refs. \cite{ISGW,ISGW2}
gives exactly Eq. (16), including the $+{3 \over 2} \epsilon$ term \cite{longinclusives}, as
expected \cite{Lipkin}. I also note that

\bigskip

1.	At $T^* \rightarrow \infty$, Eq. (16) is of the form 
$1+O(\epsilon^2)+O(\epsilon/T^*)$ as required by the OPE.

2.	There are no other terms of order $1$,
$\epsilon$, or $\epsilon T^*$ 
possible beyond those shown:  a more accurate treatment
of $\Gamma (B \rightarrow D \phi)$ could only generate
$\epsilon^2$, $\epsilon^2 T^*$, $\epsilon^2 T^{*2}$,~...
terms;  a more accurate treatment of $\Gamma (B \rightarrow D^{**} \phi)$ could only generate
$\epsilon^2 T^*$, $\epsilon^2 T^{*2}$,~... terms; and all 
higher states first make a contribution at order $\epsilon^2 T^{*2}$ or higher.  Conversely,
we note that if, for example, $\epsilon^2 T^{*2}$ terms are 
retained, they must all cancel exactly or the requirements of the OPE would be violated
as $T^* \rightarrow \infty$.

3.	As $\Delta m \equiv m_b-m_c \rightarrow 0$, 
$R_1^{D^{**}} \rightarrow 1+O({ \Lambda_{QCD}\Delta m  \over m_b^2})$ as required \cite{SVcomment}.

4.	Near $T^*_{max} \equiv m_b-m_c$, $\epsilon T^*_{max} $ is in general large.  
This observation corresponds in the usual language of heavy quark symmetry to the statement that 
the natural scale of the slope $\rho^2$
of the Isgur-Wise function is of order unity.  It is
also consistent with the experimental observation that 
$\vert \xi_{DB}  \vert^2$ has dropped to less than half its value between zero
and maximum recoil.  Given this, the extension of Eq. (16) to higher orders in 
$T^*$ will require
a ``conspiracy" of the entire spectrum of possible hadronic final states.  We may nevertheless  use
Eq. (16) across the full range of $T^*$ as an indicator of the 
$\Lambda_{QCD}/m_Q$ effects arising from the order 1 and order $T^*$ terms in the
expansion of $R$. This corresponds to a ``best case" assumption that 
duality is locally perfect for the terms $T^{*n}$
with $n>1$.

\bigskip

\noindent Thus, while extreme, this truncation has
all the properties required by the OPE and so stands as a simple explicit example of
the existence of the claimed duality-violating 
$\Lambda_{QCD}/m_Q$ effects  for finite $T^*$.

    It is straightforward to introduce a number of simple variants of this 
prototypical model.  The first corresponds to the more realistic case where
the Bjorken sum rule is only saturated by the full tower of
$s_{\ell}^{\pi_{\ell}}=1^-$ resonances so that the second term in Eq. (16) becomes
\begin{equation} 
{\epsilon \over {T^*}^{1/2}} \sum_n f_n (T^*-t^*_n)^{3/2}  \theta (T^*-t^*_n)
\end{equation}
with $\sum_n f_n=1$ and $t^*_n$ being the threshold for channel $n$.
As $T^* \rightarrow \infty$, these contributions automatically  cancel the 
$-\epsilon T^*$ term from the elastic form factor, and constrain the $O(\epsilon)$
correction to give
\begin{equation} 
R_{1+2+...}^{D^{**}} =  [1 +{3\over 2} \epsilon \bar t^* - \epsilon T^*] \theta (T^*)
+ {\epsilon \over {T^*}^{1/2}} \sum_n f_n (T^*-t^*_n)^{3/2}  \theta (T^*-t^*_n)~~~,
\end{equation}
where
\begin{equation} 
\bar t^* = \sum_n f_n t^*_n
\end{equation}
is the weighted average threshold position.  Note that since some $T^*_n$ exceed
$T^*_{max}$, $R_{1+2+...}^{D^{**}}$ cannot  heal to unity in the physical
decay region.

~
\bigskip\bigskip
~
%
%
\begin{center}
~
\epsfxsize=6.0in  \epsfbox{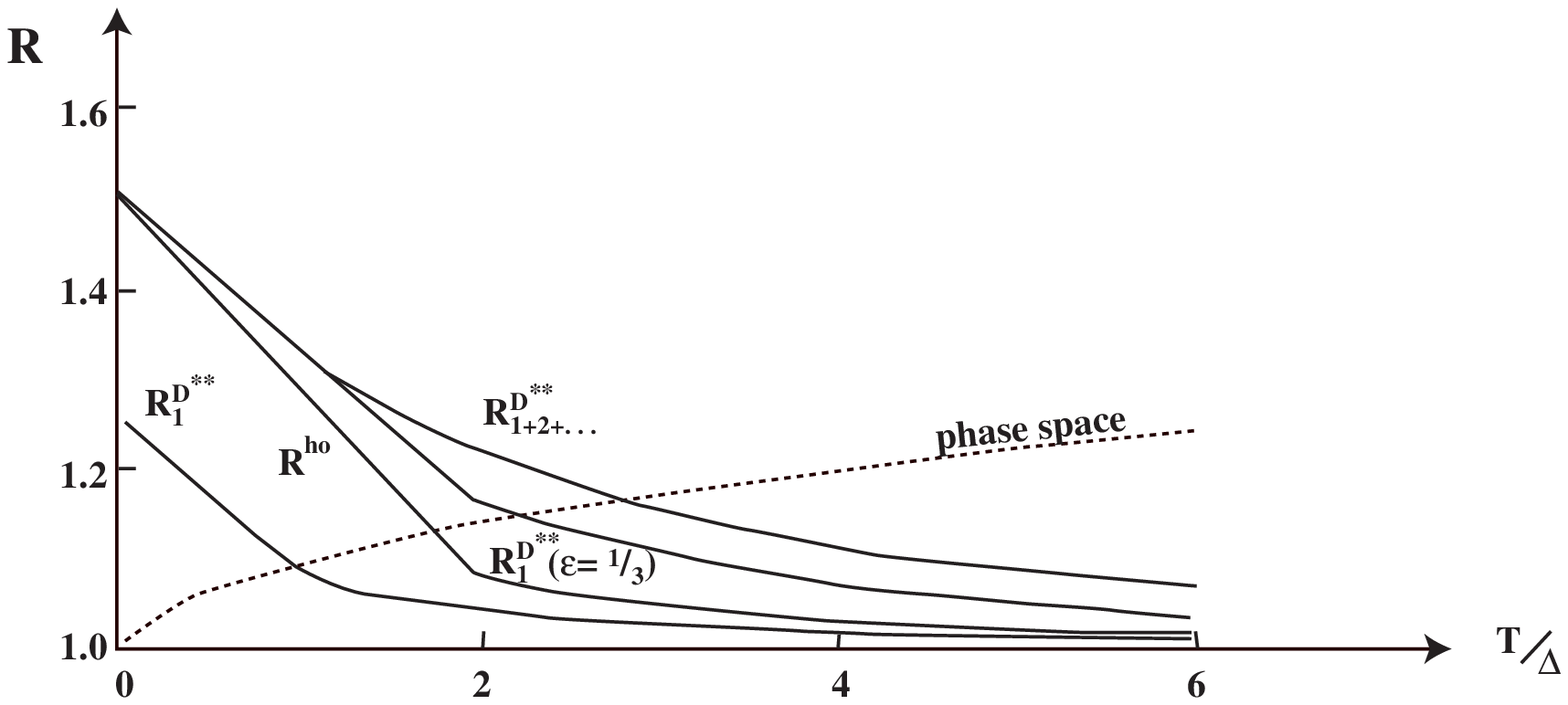}
\vspace*{0.1in}
~
\end{center}

\bigskip\bigskip

\noindent{ Fig. 4:  Four resonance models of the approach to duality:
(a) $R^{D^{**}}_1$ (with the baseline value $\epsilon=1/6$),
(b) $R^{D^{**}}_1$ (with $\epsilon=1/3$),
(c) $R^{D^{**}}_{1+2+ ...}$ (with $t^*_n=n$ and $f_n=({1 \over 2})^n$ so that $\bar t^*=2$), and
(d) $R^{ho}$. The ``baseline" parameters follow from the observations that
$\rho^2 \sim 1$, $(w-1)_{max} \sim 1/2$ and $(m_b-m_c)/\Delta \sim 6$. The alternative
$\epsilon=1/3$ corresponds to the case $T^*_{max}=3$, {\it i.e.,} to using
$\Delta_{eff}=2 \Delta$ for the mean location of the $s_{\ell}^{\pi_{\ell}}=1^-$ strength.}

\bigskip\bigskip\bigskip

	As described above, both 
$R_{1}^{D^{**}}$ and 
$R_{1+2+...}^{D^{**}}$ are still ``best case" truncations which assume exact cancellations of 
$\epsilon^2 T^*$, $\epsilon^2 T^{*2}$,~... terms.  While sufficient
for the purposes of this study, this limitation is easily removed: 
it is straightforward to recursively ``construct duality" to create models to the
required order in $\epsilon$ to any finite order in $T^*$.  Consider, 
for example, 
\begin{eqnarray} 
R^{ho} = {exp(-\epsilon T^*) \over {T^{*1/2}}} 
&&\Bigl(
[1 +{3\over 2} \epsilon]  {T^*}^{1/2} \theta (T^*)\nonumber \\
&&+\epsilon[1 +{5\over 2} \epsilon +{35 \over 16}\epsilon^2 +{35 \over 32}\epsilon^3 +{385 \over 1024}\epsilon^4 + ...]
  (T^*-1)^{3/2} \theta (T^*-1)\nonumber \\
&&+{1 \over 2!}\epsilon^2[1 +{7\over 2} \epsilon +{21 \over 4}\epsilon^2 +{77 \over 16}\epsilon^3 + ...] 
  (T^*-2)^{5/2} \theta (T^*-2)\nonumber \\
&&+{1 \over 3!}\epsilon^3[1 +{9\over 2} \epsilon +{297 \over 32}\epsilon^2 +...]  (T^*-3)^{7/2} \theta (T^*-3)\nonumber \\
&&+{1 \over 4!}\epsilon^4[1 +{11\over 2} \epsilon + ...]  (T^*-4)^{9/2} \theta (T^*-4)\nonumber \\
&&+{1 \over 5!}\epsilon^5[1+ ...] (T^*-5)^{11/2} \theta (T^*-5)\nonumber + ...
\Bigr)~~~,
\end{eqnarray}

\bigskip
\noindent where the ellipses denote terms of order $\epsilon^6 {T^*}^n$ with $1 \leq n \leq 5$ and all terms of
order $\epsilon^m {T^*}^m$ and higher with $m > 5$.  This harmonic-oscillator-like expansion is 
accurate even at $T^*=5$ up to corrections of order $\Lambda_{QCD}^2/m_Q^2$.

	   The three models just introduced are all based on the duality of $b \rightarrow c \phi$
to a simple tower of $c \bar d$ resonances controlled by the single scale $\Delta$. 
Figure 4 shows that the thresholds associated 
with such towers could easily be a source of duality-violating 
$\Lambda_{QCD}/m_Q$ corrections of order 10\% in $b \rightarrow c$
decays. This must be a cause for concern
in comparing inclusive calculations with experiment.

   I am even more alarmed by 
processes which could give a high-mass nonperturbative tail to the recoil mass distribution.
The hadronization of $b \rightarrow c \phi$ will not be
saturated by ordinary quark model $c \bar d$ states even in the large $N_c$ limit:
hybrid mesons ({\it i.e.,} states with a $c \bar d$ valence structure but with
internal gluonic excitation) will also contribute. Such states are expected at 
substantially higher masses than the ordinary
quark model states. Moreover, their production will not be exhausted until
the constituent $\bar d$ antiquark in the $D$ meson has been fully resolved into a current
quark at high recoil momentum $p_c >> 1$ GeV. For a crude estimate of 
the effects of the delayed onset of these states, I 
take a simple two-component resonance model consisting
of ``normal" $c \bar d$ resonances with $\bar t^*_{c \bar d}$  and $c \bar d$ hybrids with 
$\bar t^*_{hybrid}$ substantially larger.  If we assume that the
latter are responsible for a fraction $\kappa$ of $\rho^2_{dyn}$, then we would have
\begin{eqnarray} 
R^{hybrid} &=&  [1 +{3\over 2} \epsilon \bar t^* - \epsilon T^*] \theta (T^*) \nonumber \\
&&+ (1-\kappa)\epsilon {(T^*-\bar t^*_{c \bar d})^{3/2} \over {T^*}^{1/2}} \theta (T^*-\bar t^*_{c \bar d}) \nonumber \\
&&+ \kappa\epsilon {(T^*-\bar t^*_{hybrid})^{3/2} \over {T^*}^{1/2}} \theta (T^*-\bar t^*_{hybrid})~~~,
\end{eqnarray}
with $\bar t^*=(1-\kappa) \bar t^*_{c \bar d}+ \kappa \bar t^*_{hybrid}$.  

\bigskip

%
%
\begin{center}
~
\epsfxsize=5.0in  \epsfbox{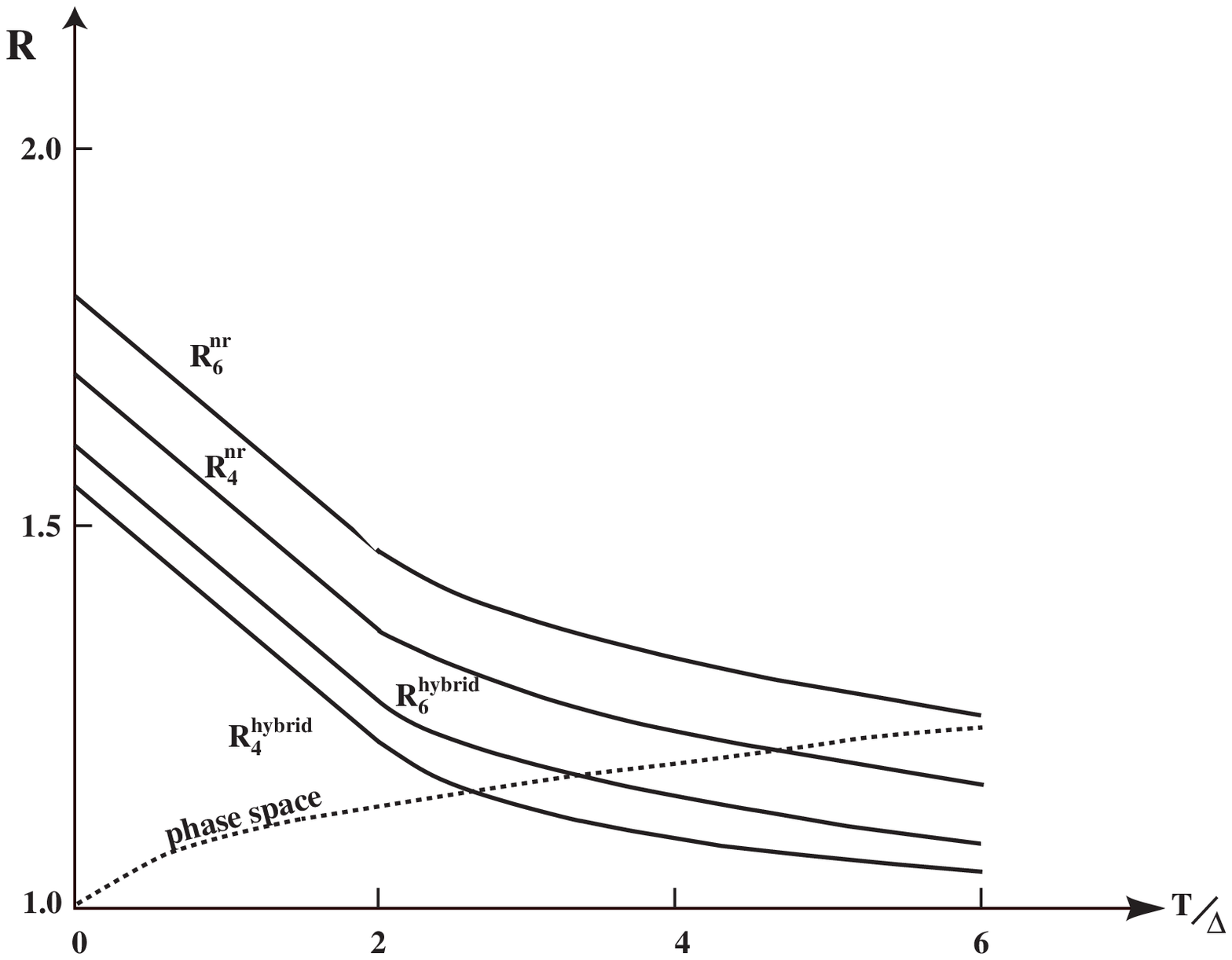}
\vspace*{0.1in}
~
\end{center}

\bigskip

\noindent{ Fig. 5: Four examples of the effects of a nonperturbative high mass tail on the recoil mass spectrum:
(a) $R_4^{hybrid}$ (with $\kappa=1/10$, $\bar t^*_{c \bar d}=2$,   and $\bar t^*_{hybrid}=4$),
(b) $R_6^{hybrid}$ (as in (a), but with $\bar t^*_{hybrid}=6$),
(c) $R_4^{nr}$ (with $\lambda=1/5$, $\bar t^*_{c \bar d}=2$,  $T^*_{min}=2$, and $s=4$), and
(d) $R_6^{nr}$ (as in (c) but with $s=6$). The values of $\kappa$ and $\lambda$ are based on 
the model-dependent estimates of Ref. \cite{exlusincluscomment}, but are certainly
reasonable ({\it e.g.,} $\lambda$ is a $1/N_c$ effect). The illustrative values for
$\bar t^*_{hybrid}$ are based on the high threshold for hybrids and their presumed
``hard" production mechanism. The nonresonant spectrum is assumed to have the form
$\rho(t^*)={1\over s}exp({T^*_{min}-t^* \over s})\theta(t^*-T^*_{min})$, with the
choices for $s$ reflecting the assumed persistence of nonresonant contributions to invariant masses
of order 2 GeV above threshold.} 

\bigskip\bigskip

	    I suspect that $1/N_c$-suppressed nonresonant contributions are an even more serious
source of delayed compensation of duality. An appropriate model for such continua would be
\begin{eqnarray} 
R^{nr} &=&  [1 +{3\over 2} \epsilon \bar t^* - \epsilon T^*] \theta (T^*) \nonumber \\
&&+ (1-\lambda)\epsilon {(T^*-\bar t^*_{c \bar d})^{3/2} \over {T^*}^{1/2}} \theta (T^*-\bar t^*_{c \bar d}) \nonumber \\
&&+ \lambda\epsilon \int_{T^*_{min}}^{T^*}dt^* \rho (t^*){(T^*-\bar t^*)^{3/2} \over {T^*}^{1/2}} ~~~,
\end{eqnarray}
where $\lambda$ is the fraction of $\rho^2_{dyn}$ due to nonresonant states 
and $\rho(t^*)$ is the appropriate normalized spectral function ($\int_{T^*_{min}}^{\infty}dt^* \rho (t^*)=1$)
which begins at $T^*_{min}$.  In this situation,
$\bar t^*=(1-\lambda) \bar t^*_{c \bar d}+ \lambda \int_{T^*_{min}}^{\infty}dt^* \rho (t^*)t^*$.
Nonperturbative quark pair creation leading to $\bar B \rightarrow X_c Y \phi$ 
may be expected \cite{nonresonant} to persist up to 2 GeV above threshold.

   Figure 5 shows that modest couplings to either hybrids or high mass continua could lead to even more
substantial duality violations than those associated with the delayed onset of the normal $c \bar d$ resonances.

   Although my main focus has been on heavy-to-heavy transitions, the  
physics issues raised here (if not their explicit forms) are also
relevant for  heavy-to-light transitions.  
Before concluding, let me therefore point out a simple application of the OPE to inclusive
heavy-to-light transitions where it seems certain to me that they will fail:  Cabibbo-forbidden charm decays. 
(Even though such decays
might be an unimportant application of the inclusive calculations in practice, they provide a valid theoretical testing
ground for their accuracy.)  In particular, consider the $c\rightarrow d \bar \ell \nu_{\ell}$ 
decays of the $D^0$ and $D^+$.  They will be dominated by 
the channels $D^0 \rightarrow \pi^- \bar \ell \nu_{\ell}$  and $\rho^- \bar \ell \nu_{\ell}$ and by 
$D^+ \rightarrow \pi^0 \bar \ell \nu_{\ell}$, $\eta \bar \ell \nu_{\ell}$, 
$\eta ' \bar \ell \nu_{\ell}$, $\rho^0 \bar \ell \nu_{\ell}$, and $\omega \bar \ell \nu_{\ell}$.  
Since the OPE corrections in the $D^0$ and $D^+$ are {\it identical}, their Cabibbo-forbidden 
semileptonic partial widths and spectral distributions are predicted to be identical.  However, simple 
isospin symmetry implies that 
$\Gamma (D^+ \rightarrow \pi^0 \bar \ell \nu_{\ell})= {1 \over 2} \Gamma(D^0 \rightarrow \pi^- \bar \ell \nu_{\ell})$, 
so the 
inclusive Cabibbo-forbidden rates can only be equal if 
$\Gamma (D^+ \rightarrow \eta \bar \ell \nu_{\ell})+\Gamma(D^+ \rightarrow \eta ' \bar \ell \nu_{\ell})
= \Gamma(D^+ \rightarrow \pi^0 \bar \ell \nu_{\ell})$.  In many models this latter relation 
would be true if $m_{\eta}=m_{\eta '}=m_{\pi}$, since it is rather
natural for the squares of matrix elements to satisfy its analogue.  However, with 
real phase space factors, this relation is typically
badly broken. Since  Cabibbo-forbidden
decays, like their Cabibbo-allowed counterparts, will receive little excited state compensation
given the available phase space,
I expect this prediction to fail.

   Finally, I note that the duality-violating effects I have 
highlighted here will have an effect on the long-standing 
$\bar B$ semileptonic branching ratio puzzle \cite{BRproblem}. 
Since the hadronic
mass distribution in $b \rightarrow c \bar u d$ is weighted toward higher
masses than the leptonic mass distribution in 
$b \rightarrow c \ell \bar \nu_{\ell}$, the ratio
of these two rates will be changed.
    
   In summary, I have shown here that hadronic thresholds lead to $\Lambda_{QCD}/m_Q$
violations of duality in $b \rightarrow c$ decays which  do not explicitly appear
in the operator product expansion. Since such violations cannot appear as the
$b \rightarrow c$ energy release $T \rightarrow \infty$,
there are ``conspiracies"  ({\it i.e.}, sum rules) which relate
hadronic thresholds and transition form factors. As emphasized by Bigi, Uraltsev, Shifman,
Vainshtein, and others \cite{Russianinclusives,otherinclusives,Lipkin}, these relations tend to compensate
the otherwise extremely large $\Lambda_{QCD}/m_Q$ effects even at small $T$. In this paper
I have displayed several models of such hadronic compensation mechanisms which
indicate that these duality-violating $\Lambda_{QCD}/m_Q$ effects could nevertheless be very substantial.
While the examples I have selected are perhaps pessimistic,
they indicate that these effects
must be better understood before inclusive methods can be 
applied with confidence to heavy quark semileptonic decays.

\bigskip\bigskip

\noindent{\bf ACKNOWLEDGEMENTS}

    I am very grateful to  number of colleagues who examined and commented upon
a draft version of this paper, including Adam Falk, Howard Georgi, Richard Lebed, 
Zoltan Ligeti, Matthias Neubert, Iain Stewart, and Mark Wise.

    I am particularly indebted to Ikaros Bigi, Misha Shifman, Nikolai Uraltsev, and Arkady Vainshtein
who replied with a detailed and very pedagogical explanation of
their understanding of the effects discussed in this paper. I
certainly learned more from this exchange than they did, and substantial errors in the draft
version were corrected as a result.

\vfill\eject


\begin{references}
\bibitem{masscomment} As argued in Refs. \cite{ISGW,ISGW2} by considering
the free recoil of the final state quark $c$ against a static spectator $\bar d$ quark, $\Delta m_{X_c}$
should be roughly $[(m_c+m_d)^2+m_dm_b^{-1}(m_b-m_c)^2]^{1\over 2}-(m_c+m_d) \simeq  0.2$ GeV, 
consistent with the data and the inclusive models shown in 
Figure 1, and certainly not leading to a small expansion parameter $\Lambda_{QCD}/
\Delta m_{X_c}$ for $b \rightarrow c$ decays.
\end{references}
\end{document}